\documentclass[aps, a4paper, floatfix, longbibliography, twocolumn,superscriptaddress,nofootinbib,numbers,sort&compress]{revtex4-1}
\usepackage[utf8]{inputenc}
\usepackage{epsfig}
\usepackage[T1]{fontenc}
\usepackage[english]{babel}
\usepackage[table]{xcolor}
\usepackage{t1enc}
\usepackage{graphicx}
\usepackage{amssymb}
\usepackage{amsmath}
\usepackage{relsize}
\usepackage{overpic}
\usepackage{bm}
\usepackage{hyperref}

\usepackage[normalem]{ulem}

\newcommand{\dd}{\mathrm{d}}

\usepackage{mathtools}
\def\multiset#1#2{\ensuremath{\left(\kern-.3em\left(\genfrac{}{}{0pt}{}{#1}{#2}\right)\kern-.3em\right)}}

\usepackage{amsmath}

\usepackage{verbatim}
\usepackage{overpic}
\usepackage{booktabs}
\usepackage{placeins}

\newcommand{\boldlabel}[1]{\textsf{\textbf{#1}}}
\usepackage{etoolbox}
\patchcmd{\section}
  {\centering}
  {\raggedright}
  {}
  {}
 \patchcmd{\section}
  {\normalfont}
  {\sffamily}
  {}
  {}
\patchcmd{\subsection}
  {\centering}
  {\raggedright}
  {}
  {}
 \patchcmd{\subsection}
  {\normalfont}
  {\sffamily}
  {}
  {}

\begin{document}

\title{\sffamily Modeling sequences and temporal networks with dynamic community structures}

\author{Tiago P. Peixoto}
\email{t.peixoto@bath.ac.uk}
\affiliation{Department of Mathematical Sciences and Centre for Networks
and Collective Behaviour, University of Bath, Claverton Down, Bath BA2
7AY, United Kingdom}
\affiliation{ISI Foundation, Via Alassio 11/c, 10126 Torino, Italy}

\author{Martin Rosvall}
\affiliation{Integrated Science Lab, Department of Physics, Ume\r{a} University, SE-901 87 Ume\r{a}, Sweden}


\begin{abstract} 
In evolving complex systems such as air traffic and social
organizations, collective effects emerge from their many components'
dynamic interactions. While the dynamic interactions can be
represented by temporal networks with nodes and links that change over
time, they remain highly complex. It is therefore often necessary to
use methods that extract the temporal networks' large-scale dynamic community
structure. However, such methods are subject to overfitting or suffer
from effects of arbitrary, a priori imposed timescales, which should
instead be extracted from data. Here we simultaneously address both
problems and develop a principled data-driven method that determines
relevant timescales and identifies patterns of dynamics that take
place on networks as well as shape the networks themselves. We base
our method on an arbitrary-order Markov chain model with community
structure, and develop a nonparametric Bayesian inference framework
that identifies the simplest such model that can explain temporal
interaction data.
\end{abstract}

\maketitle

\section*{Introduction}

To reveal the mechanisms of complex systems, researchers identify
large-scale patterns in their networks of interactions with
community-detection methods~\cite{fortunato_community_2010}.
Traditionally, these methods describe only static network structures
without taking into account the dynamics that take place {on} the
networks, such as people travelling by air, or the dynamics {of}
the networks themselves, such as new routes in air traffic networks.
While the dynamics on and of networks contain crucial information about
the systems they represent, only recently have researchers showed how to
incorporate higher-order Markov chains to describe dynamics on networks~\cite{
rosvall_memory_2014,lambiotte_effect_2015,de_domenico_identifying_2015,
scholtes2017network} and higher-order temporal
structures to describe dynamics of networks~\cite{holme_modern_2015,mucha_community_2010,rosvall_mapping_2010,
bassett_robust_2013, pfitzner_betweenness_2013, bazzi2016community,
sarzynska2016null, gauvin_detecting_2014,
xu_dynamic_2013, peixoto_inferring_2015, macmahon_community_2015,
ghasemian2016detectability, zhang_random_2016}.
However, both
avenues of research have encountered central limitations:
First, methods that use higher-order memory to describe dynamics on networks rely on
extrinsic methods to detect the appropriate memory
order~\cite{rosvall_memory_2014,xu_representing_2016}.
Second, methods that attempt to describe dynamics of networks adapt
static descriptions by aggregating time windows into discrete
layers~\cite{kivela_multilayer_2014,xu_dynamic_2013, xu_dynamic_2014, 
xu2015stochastic,gauvin_detecting_2014,peixoto_inferring_2015,
ghasemian2016detectability}, and ignore dynamics within the time windows.
Thus, both methods for dynamics on and of networks require or
impose ad hoc timescales that can obscure essential dynamic community structure.

Furthermore, when trying to determine the timescales solely from data,
the curse of dimensionality strikes: the large number of degrees of
freedom makes the higher-order descriptions prone to {overfitting}
when random fluctuations in high-dimensional data are mistaken for
actual structure~\cite{guimera_modularity_2004}.  Without a principled
method with effective model selection to counteract this increasing
complexity, it becomes difficult to separate meaningful dynamic
community structure from artefacts.

\begin{figure*}
  \centering
  \begin{tabular}{cc}
    \begin{minipage}{.5\textwidth}
  \includegraphics[width=.72\columnwidth,trim=0 0 0 0]{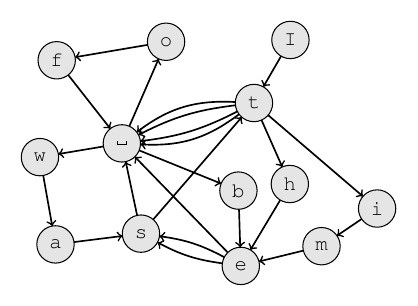}
  \put(-200,120){\larger\boldlabel{a}}
  \end{minipage} &
  \begin{minipage}{.5\textwidth}
  \includegraphics[width=1\columnwidth,trim=0 0 0 0]{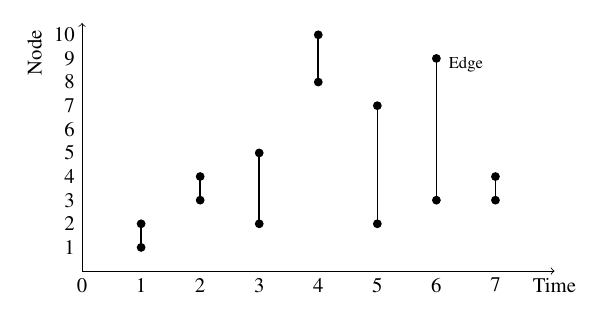}
  \put(-270,120){\larger\boldlabel{b}}
  \end{minipage}
  \end{tabular}

  \caption{{Unified modelling of dynamics on and of networks.}
  Our modeling framework simultaneously describes:
  (\boldlabel{a}) Arbitrary dynamics taking place on networks, represented as
  a sequence of arbitrary {tokens} that are associated with nodes, in this
  example $\{x_t\}=\text{\texttt{"It␣was␣the␣best␣of␣times"}}$. (\boldlabel{b})
  Dynamics of networks themselves, where the tokens are node-pair edges that appear in sequence, in this example $\{x_t\}=\{(1, 2), (4,
  3), (5, 2), (10, 8), (7, 2), (9, 3), (3, 4)\}$. \label{fig:diagram}}
\end{figure*}

To overcome these model selection and arbitrary timescale problems, we
present a general and principled data-driven method by simultaneously tackling dynamics
on and of networks (see Fig.~\ref{fig:diagram}).  In
contrast to approaches that incorporate temporal layers in methods for
static network descriptions, we build our approach on describing the
actual dynamics.  We first formulate a generative model of discrete
temporal processes based on arbitrary-order Markov chains with community
structure~\cite{baum_maximization_1970,rabiner_introduction_1986,jaaskinen_sparse_2014,xiong_recursive_2016}. Since
our model generates event sequences, it does not aggregate data
in time
windows~\cite{xu_dynamic_2013,gauvin_detecting_2014,peixoto_inferring_2015,ghasemian2016detectability},
and, other than the Markov model assumption, needs no a priori imposed timescales.
This model can be used to describe dynamics taking place on network
systems that take into account higher-order memory
effects~\cite{rosvall_memory_2014,lambiotte_effect_2015} of arbitrary
order. We then use the model to describe
temporal networks, where the event sequence represents the
occurrence of edges in the network~\cite{pfitzner_betweenness_2013}.

In both cases, we employ a nonparametric Bayesian inference framework
that allows us to select, according to the statistical evidence
available, the most parsimonious model among all its variations. Hence
we can, for example, identify the most appropriate Markov order and the
number of communities without overfitting. In particular, if the
dynamics on or of a network are random, our method will not identify
any spurious patterns from noise but conclude that the data lack
structure. As we also show, the model can be used to predict future
network dynamics and evolution from past observations. Moreover, we
provide publicly available and scalable code with log-linear
complexity in the number of nodes independent of the number of groups.
\newpage


\section*{Results}

\subsection*{Inference of Markov chains}\label{sec:Markov}

Here we consider general time-series composed of a sequence of
discrete observations $\{x_t\}$, where $x_t$ is a single {token}
from an alphabet of size $N$ observed at discrete time $t$, and
$\bm{x}_{t-1} = (x_{t-1}, \dots, x_{t-n})$ is the {memory} of
the previous $n$ tokens at time $t$ (see Fig.~\ref{fig:diagram}). An
$n$th-order Markov chain with transition probabilities
$p(x_t|\bm{x}_{t-1})$ generates such a sequence with probability
\begin{equation}\label{eq:Markov}
  P(\{x_t\} | p) = \prod_tp(x_t|\bm{x}_{t-1})
                 = \prod_{x,\bm{x}}p(x|\bm{x})^{a_{x,\bm{x}}},
\end{equation}
where $a_{x,\bm{x}}$ is the number of transitions $\bm{x} \to x$ in
$\{x_t\}$. Given a
specific sequence $\{x_t\}$, we want to infer the
transitions probabilities $p(x|\bm{x})$. The simplest approach is to compute the maximum-likelihood estimate, that is\
\begin{equation}
  \hat{p}(x|\bm{x}) = \underset{p(x|\bm{x})}{\operatorname{argmax}}\, P(\{x_t\} | p) = \frac{a_{x,\bm{x}}}{a_{\bm{x}}},
\end{equation}
where $a_{\bm{x}} = \sum_x a_{x,\bm{x}}$, which amounts simply to the
frequency of observed transitions. Putting this back into the likelihood
of Eq.~\ref{eq:Markov}, we have
\begin{equation}\label{eq:mle}
  \ln P(\{x_t\} | \hat{p}) = \sum_{x,\bm{x}}a_{x,\bm{x}} \ln \frac{a_{x,\bm{x}}}{a_{x}}.
\end{equation}
This can be expressed through the total number of observed transitions $E =
\sum_{x,\bm{x}} a_{x,\bm{x}}$ and the conditional entropy
$H(X|\bm{X}) = -\sum_{\bm{x}}\hat{p}(\bm{x}) \sum_x
\hat{p}(x|\bm{x}) \ln \hat{p}(x|\bm{x})$ as $\ln P(\{x_t\} | \hat{p})
= -EH(X|\bm{X})$.  Hence, the maximisation of the likelihood in
Eq.~\ref{eq:Markov} yields the transition probabilities
that {most compress} the sequence. There is, however, an important
caveat with this approach. It cannot be used when we are
interested in determining the most appropriate Markov order $n$ of the
model, because the maximum likelihood in
Eq.~\ref{eq:mle} increases with $n$. In general, increasing number of
memories at fixed number of transitions leads to decreased conditional entropy.
Hence, for some large enough value of
$n$ there will be only one observed transition conditioned on every
memory, yielding a zero conditional entropy and a maximum likelihood of
1. This would be an extreme case of {overfitting}, where by
increasing the number of degrees of freedom of the model it is impossible
to distinguish actual structure from stochastic fluctuations. Also, this
approach does not yield true compression of the data, since it does not
describe the increasing model complexity for larger values of $n$, and
thus is crucially incomplete. To address this problem, we use a Bayesian
formulation, and maximise instead the complete evidence
\begin{equation}\label{eq:bayes}
  P(\{x_t\}) =  \int P(\{x_t\} | p)P(p)\,\dd p,
\end{equation}
which is the sum of all possible models weighted according to
prior probabilities $P(p)$ that encode our a priori
assumptions. This approach gives the
correct model order for data sampled from Markov chains as long as
there are enough statistics that balances the structure present in the data
with its statistical weight, as well as meaningful values when
this is not the case~\cite{strelioff_inferring_2007}.

Although this Bayesian approach satisfactorily addresses the overfitting
problem, it misses opportunities of detecting large-scale structures in data. As we
show below, it is possible to extend this model in such a way as to make
a direct connection to the problem of finding communities in networks,
yielding a stronger explanatory power when modelling sequences, and
serving as a basis for a model where the sequence itself represents a
temporal network.

\subsection*{Markov chains with communities}\label{sec:markov_communities}

\begin{figure*}
  \begin{tabular}{rr}
    \begin{overpic}[width=.472\textwidth,trim=0 -1.6cm 0 0]{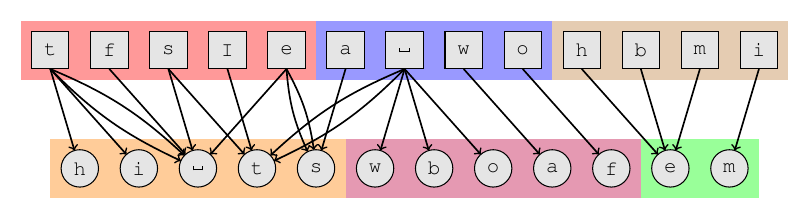}
      \put(0,43){\larger\boldlabel{a}}
    \end{overpic}
    &
    \begin{overpic}[width=.32\textwidth]{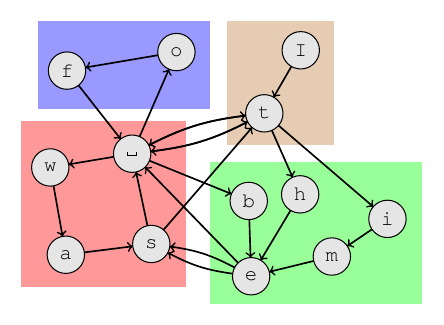}
      \put(84,65){\larger\boldlabel{b}}
    \end{overpic}
    \\
    \multicolumn{2}{r}{\begin{overpic}[width=.8\textwidth]{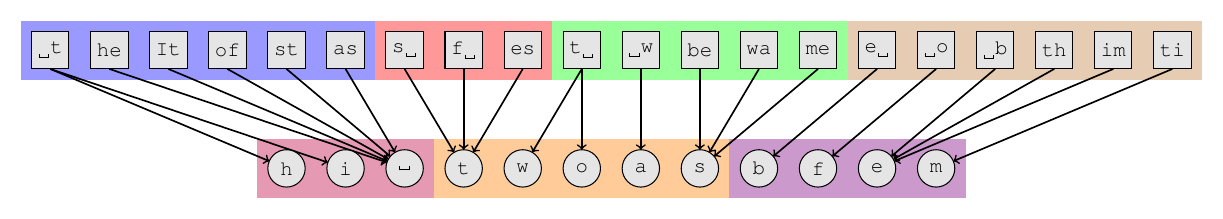}\put(0,0){\larger\boldlabel{c}}\end{overpic}}
    \end{tabular}

    \caption{{Schematic representation of the Markov model
    with communities. The token sequence
    $\{x_t\}=\text{\texttt{"It␣was␣the␣best␣of␣times"}}$} represented with nodes for memories (top row) and tokens (bottom row), and with directed
    edges for transitions in different variations of the model. (\boldlabel{a}) A partition of the tokens
    and memories for an $n=1$ model. (\boldlabel{b}) A unified formulation of an $n=1$
    model, where the tokens and memories have the same partition, and
    hence can be represented as a single set of nodes. (\boldlabel{c}) A partition
    of the tokens and memories for an $n=2$ model. \label{fig:model}}
\end{figure*}

Instead of directly inferring the transition probabilities of
Eq.~\ref{eq:Markov}, we propose an alternative formulation: We assume
that both memories and tokens are distributed in disjoint groups
(see Fig.~\ref{fig:model}). That is, $b_{x} \in [1, 2, \ldots , B_\text{N}]$ and
$b_{\bm{x}} \in [B_\text{N} + 1, B_\text{N} + 2, \ldots, B_\text{N}
+ B_{\text{M}}]$ are the group memberships of the tokens and memories uniquely assigned in $B_\text{N}$ and $B_{\text{M}}$ groups,
respectively, such that the transition probabilities can be parametrised
as
\begin{equation}\label{eq:block_prob}
  p(x|\bm{x}) = \theta_x\lambda_{b_xb_{\bm{x}}}.
\end{equation}
Here $\theta_x$ is the relative probability at which token $x$ is
selected among those that belong to the same group, and $\lambda_{rs}$ is
the overall transition probability from memory group $s=b_{\bm{x}}$
to token group $r=b_x$. The parameter $\theta_x$ plays an analogous
role to degree-correction in the SBM~\cite{karrer_stochastic_2011},
and is together with the Bayesian description the main difference from
the sparse Markov chains developed in
Refs.~\cite{jaaskinen_sparse_2014,xiong_recursive_2016}.
In the case $n=1$, for example, each token appears twice in the
model, both as token and memory. An alternative and often useful
approach for $n=1$ is to consider a single unified partition for both
tokens and memories, as shown in Fig.~\ref{fig:model}b and described in
detail in the Methods section The unified first-order model.
In any case, the maximum likelihood parameter estimates are
\begin{equation}
  \hat{\lambda}_{rs} =\frac{e_{rs}}{e_s}, \quad \hat{\theta}_x = \frac{k_x}{e_{b_x}},
\end{equation}
where $e_{rs}$ is the number of observed transitions from group $s$ to
$r$, $e_s = \sum_te_{ts}$ is the total outgoing transitions from group
$s$ if $s$ is a memory group, or the total incoming transition if it
is a token group. The labels $r$ and $s$ are used indistinguishably to
denote memory and token groups, since it is only their numerical value
that determines their kind. Finally, $k_x$ is the total number of
occurrences of token $x$. Putting this back in the likelihood, we have
\begin{equation}\label{eq:sbm_ml}
  \ln\hat{P}(\{x_t\} | b, \hat{\lambda}, \hat{\theta}) = \sum_{r<s}e_{rs} \ln
  \frac{e_{rs}}{e_re_s} + \sum_xk_x\ln k_x.
\end{equation}
This is almost the same as the maximum likelihood of the
degree-corrected stochastic block model
(DCSBM)~\cite{karrer_stochastic_2011}, where $a_{x,\bm{x}}$ plays the
role of the adjacency matrix of a bipartite multigraph connecting tokens
and memories. The only differences are constant terms that do not alter
the position of the maximum with respect to the node partition. This
implies that for undirected networks without higher-order memory,
there is no difference between inferring the structure directly from
its topology or from dynamical processes taking place on it, as we show
in detail in the Methods section Equivalence between structure and dynamics.

\begin{figure}[t!]
  \includegraphics[width=.95\columnwidth]{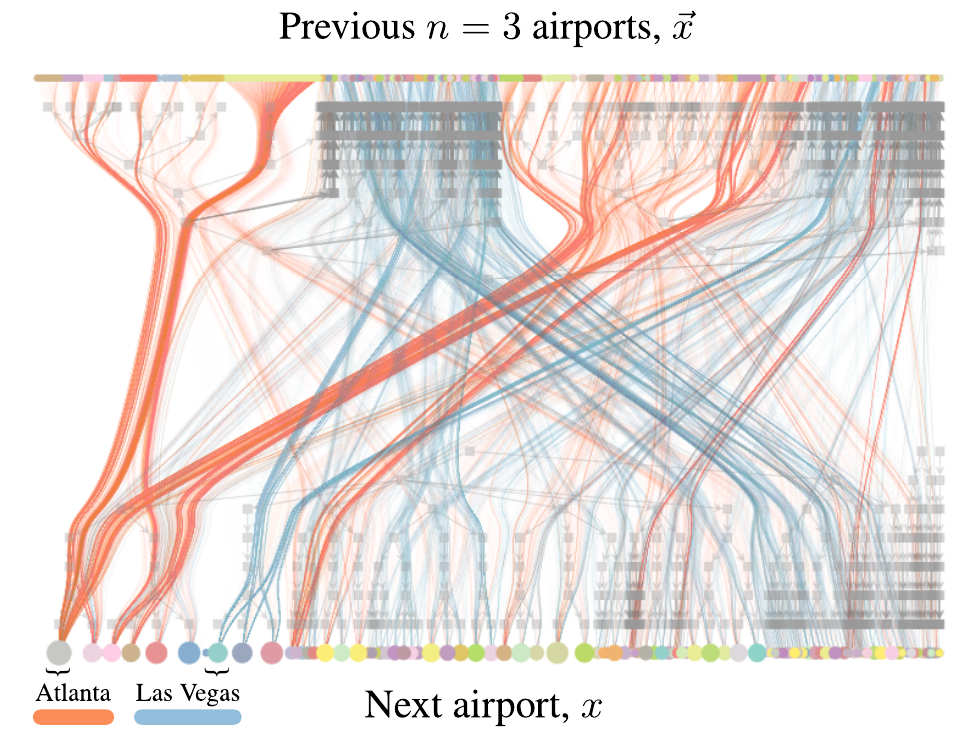} \caption{{
  Selection of US flight itineraries for a third-order model.}
  The itineraries contain stops in Atlanta or Las Vegas.
  Edges incident on memories
  of the type $\bm{x}=(x_{t-1},\text{Atlanta},x_{t-3})$ in
  red and $\bm{x}=(x_{t-1},\text{Las Vegas},x_{t-3})$ 
  in blue. The node colours and overlaid hierarchical division
  derive from the $n=3$ model inferred for the whole
  dataset. \label{fig:air2011}}
\end{figure}

\begin{table*}
  \resizebox{\textwidth}{!}{
  \begin{tabular}{r|rrrr|rrrr|rrrr|rrrr}
    &\multicolumn{4}{c|}{US flight itineraries} & \multicolumn{4}{c|}{War and Peace} & \multicolumn{4}{c|}{Taxi movements} & \multicolumn{4}{c}{RockYou password list} \\ \hline \hline
    $n$
    & $B_\text{N}$ & $B_{\text{M}}$ & \multicolumn{1}{c}{$\Sigma$} &\multicolumn{1}{c|}{$\Sigma'$}
    & $B_\text{N}$ & $B_{\text{M}}$ & \multicolumn{1}{c}{$\Sigma$} &\multicolumn{1}{c|}{$\Sigma'$}
    & $B_\text{N}$ & $B_{\text{M}}$ & \multicolumn{1}{c}{$\Sigma$} &\multicolumn{1}{c|}{$\Sigma'$}
    & $B_\text{N}$ & $B_{\text{M}}$ & \multicolumn{1}{c}{$\Sigma$} &\multicolumn{1}{c|}{$\Sigma'$}
    \\ \hline
        $1$ & $384$& $365$& $364,385,780$& $365,211,460$& $65$& $71$& $11,422,564$& $11,438,753$& $387$& $385$& $2,635,789$& $2,975,299$\cellcolor[gray]{0.8}& $140$& $147$& $1,060,272,230$& $1,060,385,582$\\
    $2$ & $386$& $7605$& $319,851,871$& $326,511,545$\cellcolor[gray]{0.8}& $62$& $435$& $9,175,833$& $9,370,379$& $397$& $1127$& $2,554,662$\cellcolor[gray]{0.8}& $3,258,586$& $109$& $1597$& $984,697,401$& $987,185,890$\\
    $3$ & $183$& $2455$& $318,380,106$\cellcolor[gray]{0.8}& $339,898,057$& $70$& $1366$& $7,609,366$& $8,493,211$\cellcolor[gray]{0.8}& $393$& $1036$& $2,590,811$& $3,258,586$& $114$& $4703$& $910,330,062$& $930,926,370$\cellcolor[gray]{0.8}\\
    $4$ & $292$& $1558$& $318,842,968$& $337,988,629$& $72$& $1150$& $7,574,332$\cellcolor[gray]{0.8}& $9,282,611$& $397$& $1071$& $2,628,813$& $3,258,586$& $114$& $5856$& $889,006,060$\cellcolor[gray]{0.8}& $940,991,463$\\
    $5$ & $297$& $1573$& $335,874,766$& $338,442,011$& $71$& $882$& $10,181,047$& $10,992,795$& $395$& $1095$& $2,664,990$& $3,258,586$& $99$& $6430$& $1,000,410,410$& $1,005,057,233$\\

    \hline
    gzip & \multicolumn{4}{c|}{$573,452,240$} & \multicolumn{4}{c|}{$9,594,000$} & \multicolumn{4}{c|}{$4,289,888$} & \multicolumn{4}{c}{$1,315,388,208$}\\
    LZMA & \multicolumn{4}{c|}{$402,125,144$} & \multicolumn{4}{c|}{$7,420,464$} & \multicolumn{4}{c|}{$2,902,904$} & \multicolumn{4}{c}{$1,097,012,288$}
  \end{tabular}}

  \caption{{Summary of inference results for empirical
  sequences.}
    Description length $\Sigma = -\log_2 P(\{x_t\},b)$ in bits,
  as well as inferred number of token groups $B_\text{N}$ and memory groups
  $B_{\text{M}}$ for different data sets and Markov
  order $n$ (for detailed descriptions, see Methods section Datasets). The
  value $\Sigma' = -\log_2 P(\{x_t\})$ corresponds to the direct
  Bayesian parametrisation of Markov chains of
  Ref.~\cite{strelioff_inferring_2007}, with noninformative
  priors. Values in grey correspond to the minimum of each column.
  The bottom rows show the compression obtained with gzip and LZMA, two
  popular variations of
  Lempel-Ziv~\cite{ziv_universal_1977,ziv_compression_1978}.\label{tab:dl_seq}}
\end{table*}

As before, this maximum likelihood approach cannot be used if we do not
know the order of the Markov chain, otherwise it will overfit. In fact,
this problem is now aggravated by the larger number of model parameters.
Therefore, we employ a Bayesian formulation and construct a
generative process for the model parameters themselves.  We do this by
introducing prior probability densities for the parameters
$\mathcal{D}_r(\{\theta_x\}|\alpha)$ and
$\mathcal{D}_s(\{\lambda_{rs}\}|\beta)$ for tokens and memories,
respectively, with hyperparameter sets $\alpha$ and $\beta$, and
computing the integrated likelihood
\begin{multline}
  P(\{x_t\}|\alpha,\beta,b) = \int \dd\theta \dd\lambda\; P(\{x_t\} |b,\lambda,\theta) \\
  \times\prod_r\mathcal{D}_r(\{\theta_x\}|\alpha)\prod_s\mathcal{D}_s(\{\lambda_{rs}\}|\beta).
\end{multline}
where we used $b$ as a shorthand for $\{b_x\}$ and $\{b_{\bm{x}}\}$.
Now, instead of inferring the hyperparameters, we can make a
noninformative choice for $\alpha$ and $\beta$ that reflects our a
priori lack of preference towards any particular
model~\cite{jaynes_probability_2003}. Doing so in this case yields a
likelihood (for details, see Methods section Bayesian Markov chains with communities),
\begin{multline}\label{eq:chain_dl}
  P(\{x_t\}|b, \{e_s\}) =  P(\{x_t\}|b,\{e_{rs}\}, \{k_x\}) \\
  \times P(\{k_x\}|\{e_{rs}\},b)P(\{e_{rs}\}|\{e_s\}),
\end{multline}
where $P(\{x_t\}|b,\{e_{rs}\}, \{k_x\})$ corresponds to the likelihood
of the sequence $\{x_t\}$ conditioned on the transitions counts
$\{e_{rs}\}$ and token frequencies $\{k_x\}$, and the remaining terms
are the prior probabilities on the discrete parameters $\{e_{rs}\}$
and $\{k_x\}$. Since the likelihood above still is conditioned on the partitions
$\{b_x\}$ and $\{b_{\bm{x}}\}$, as well as the memory group counts
$\{e_s\}$, we need to include prior probabilities on these as well to
make the approach fully nonparametric. Doing so yields a joint
likelihood for both the sequence and the model parameters,
\begin{equation}\label{eq:joint_complete}
  P(\{x_t\}, b, \{e_s\}) = P(\{x_t\}|b,\{e_s\})P(b)P(\{e_s\}).
\end{equation}
It is now possible to understand why maximizing this joint likelihood
will prevent overfitting the data. If we take its negative logarithm, it
can be written as
\begin{align}
  \Sigma &= -\log_2 P(\{x_t\}, b, \{e_s\}) \\
         &= -\log_2 P(\{x_t\}| b, \{e_s\}) - \log_2 P(b,\{e_s\})\label{eq:dl}.
\end{align}
The quantity $\Sigma$ is called the description length of the
data~\cite{grunwald_minimum_2007, rosvall_information-theoretic_2007}.
It corresponds to the amount of information necessary to describe both
the data and the model simultaneously, corresponding to the first and
second terms in Eq.~\ref{eq:dl}, respectively. As the model becomes more
complex---either by increasing the number of groups or the order of
the Markov chain---this will decrease the first term as the data
likelihood increases, but it will simultaneously increase the second
term, as the model likelihood decreases. The second term then acts as a
penalty to the model likelihood, forcing a balance between model
complexity and quality of fit. Unlike approximative penalty approaches
based solely on the number of free parameters such as
BIC~\cite{schwarz_estimating_1978} and AIC~\cite{akaike_new_1974}, which
are not to valid for network models~\cite{yan_model_2014}, the
description length of the model is exact and fully captures its
flexibility.  Because of the complete character of the description
length, minimizing it indeed amounts to achieving true compression of
data, differently from the parametric maximum likelihood approach
mentioned earlier.  Because the whole process is functionally equivalent
to inferring the SBM for networks, we can use the same
algorithms~\cite{peixoto_efficient_2014} (for a details about the
inference method, see Methods section Bayesian Markov chains with communities).

Before we continue, we point out that the selection of priors in
Eq.~\ref{eq:chain_dl} needs to be done carefully to avoid
{underfitting} the data. This happens when strong prior assumptions
obscure structures in the data~\cite{peixoto_parsimonious_2013}. We
tackle this by using hierarchical priors, where the parameter themselves
are modelled by parametric distributions, which in turn contain more
parameters, and so
on~\cite{peixoto_hierarchical_2014,peixoto_nonparametric_2017}. Besides
alleviating the underfitting problem, this allows us to represent the
data in multiple scales by a hierarchical partition of the token and
memories. We describe this in more detail in the Methods section Bayesian Markov chains with communities.

This Markov chain model with communities succeeds in providing a better
description for a variety of empirical sequences when compared with the
common Markov chain parametrisation (see Table~\ref{tab:dl_seq}). Not
only do we systematically observe a smaller description length, but we
also find evidence for higher order memory in all examples. We emphasise
that we are protected against overfitting: If we randomly shuffle the
order of the tokens in each dataset, with dominating probability we
infer a fully random model with $n=1$ and $B_\text{N}=B_{\text{M}}=1$,
which is equivalent to an $n=0$ memoryless model. 
We have verified that we infer this  model for all
analysed datasets.
Accordingly, we are not susceptible to the spurious
results of nonstatistical methods~\cite{guimera_modularity_2004}.

To illustrate the effects of community structure on the Markov dynamics,
we use the US flight itineraries as an example. In this dataset, the itineraries of
$1,272,696$ passengers were recorded, and we treat each airport stop
as a token in a sequence (for more
details, see Methods section Datasets). When we infer our model, the itinerary memories are grouped
together if their destination probabilities are similar. As a result, it
becomes possible, for example, to distinguish transit hubs from
destination hubs~\cite{rosvall_memory_2014}. We use Atlanta and Las
Vegas to illustrate: Many roundtrip routes transit through Atlanta from
the origin to the final destination and return to it two legs later on
the way back to the origin. On the other hand, Las Vegas often is the
final destination of a roundtrip such that the stop two legs later
represents a more diverse set of origins (Fig.~\ref{fig:air2011}).
Resembling the results of the map equation for network flows with memory~\cite{rosvall_memory_2014},
this pattern is captured in our model by the larger number of memory groups
that involve Las Vegas than those that involve Atlanta.
Moreover, the division between transit and destinations
propagates all the way to the upper hierarchical levels of the memory
partition.

In addition to this itinerary memory clustering, the co-clustering with
airport tokens also divides the airports into hierarchical
categories. For example, Atlanta is grouped with nearby Charlotte at the
first hierarchy level, and with Detroit, Minneapolis, Dallas and Chicago
at the third level. This extra information tells us that these airports
serve as alternative destinations to itineraries that are similar to
those that go through Atlanta. Likewise, Las Vegas is grouped together
with alternative destinations Phoenix and Denver.  This type of
similarity between airports---which is not merely a reflection of the
memory patterns---is not expressed with the assortative approach of
the map equation, which solely clusters densely connected memories with
long flow persistence times~\cite{rosvall_memory_2014}. A more direct comparison between our Bayesian inference
framework and the map equation is not meaningful, since these two
approaches represent the network divisions differently
(for a detailed discussion, see
Methods section Comparison with the map equation for network flows with memory). Indeed, it is the simultaneous
division of memories and tokens that effectively reduce the overall
complexity of the data, and provide better compression at higher
memory order. Consequently, the community-based Markov model can capture
patterns of higher-order memory that conventional methods obscure.

\subsection*{Temporal networks}\label{sec:temporal}
\begin{table*}
  \resizebox{\textwidth}{!}{
    \begin{tabular}{r|rrrrr|rrrrr|rrrrr}
&\multicolumn{5}{c|}{High school proximity ($N=327, E=5,818$)}&\multicolumn{5}{c|}{Enron email ($N=87,273$, $E=1,148,072$)}&\multicolumn{5}{c|}{Internet AS ($N=53,387$, $E=500,106$)}\\\hline\hline
$n$
& $C$ & $B_\text{N}$ & $B_{\text{M}}$ & \multicolumn{1}{c}{$\Sigma$} & \multicolumn{1}{c|}{$-\Delta \Sigma$}
& $C$ & $B_\text{N}$ & $B_{\text{M}}$ & \multicolumn{1}{c}{$\Sigma$} & \multicolumn{1}{c|}{$-\Delta \Sigma$}
& $C$ & $B_\text{N}$ & $B_{\text{M}}$ & \multicolumn{1}{c}{$\Sigma$} & \multicolumn{1}{c|}{$-\Delta \Sigma$}
\\\hline
   $0$ & $10$& --- & ---& $89,577$& $-64,129$& $1,447$& --- & ---& $19,701,405$& $-11,631,987$& $187$& --- & ---& $19,701,403$& $-8,094,541$\\
   $1$ & $10$& $9$& $9$& $82,635$\cellcolor[gray]{0.8}& $-49,216$\cellcolor[gray]{0.8}& $1,596$& $2,219$& $2,201$& $13,107,399$\cellcolor[gray]{0.8}& $-8,012,378$\cellcolor[gray]{0.8}& $185$& $131$& $131$& $10,589,136$\cellcolor[gray]{0.8}& $-6,729,923$\cellcolor[gray]{0.8}\\
   $2$ & $10$& $6$& $6$& $86,249$& $-49,533$& $324$& $366$& $313$& $16,247,904$& $-8,370,876$& $132$& $75$& $43$& $14,199,548$& $-6,921,032$\\
   $3$ &  $9$& $6$& $6$& $103,453$& $-49,746$& $363$& $333$& $289$& $26,230,928$& $-14,197,057$& $180$& $87$& $79$& $22,821,016$& $-8,133,665$\\
\hline\multicolumn{16}{c}{}\\&\multicolumn{5}{c|}{APS citations ($N=425,760$, $E=4,262,443$)}&\multicolumn{5}{c|}{\texttt{prosper.com} loans ($N=89,269$, $E=3,394,979$)}&\multicolumn{5}{c|}{Chess moves ($N=76$, $E=3,130,166$)}\\\hline\hline   $0$ & $3,774$& --- & ---& $131,931,579$& $-93,802,176$& $318$& --- & ---& $96,200,002$& $-64,428,332$& $72$& --- & ---& $66,172,128$& $-34,193,040$\\
   $1$ & $4,426$& $6,853$& $6,982$& $94,523,280$\cellcolor[gray]{0.8}& $-56,059,700$\cellcolor[gray]{0.8}& $267$& $1039$& $1041$& $59,787,374$\cellcolor[gray]{0.8}& $-30,487,941$\cellcolor[gray]{0.8}& $72$& $339$& $339$& $58,350,128$& $-30,271,323$\\
   $2$ & $4,268$& $710$& $631$& $144,887,083$& $-,100,264,678$& $205$& $619$& $367$& $109,041,487$& $-54,211,919$& $72$& $230$& $266$& $58,073,342$\cellcolor[gray]{0.8}& $-30,110,657$\cellcolor[gray]{0.8}\\
   $3$ & $4,268$& $454$& $332$& $228,379,667$& $-,120,180,052$& $260$& $273$& $165$& $175,269,743$& $-54,655,474$& $72$& $200$& $205$& $76,465,862$& $-32,120,845$\\
\hline\multicolumn{16}{c}{}\\&\multicolumn{5}{c|}{Hospital contacts ($N=75$, $E=32,424$)}&\multicolumn{5}{c|}{Infectious Sociopatterns ($N=10,972$, $E=415,912$)}&\multicolumn{5}{c|}{Reality Mining ($N=96$, $E=1,086,404$)}\\\hline\hline   $0$ & $68$& --- & ---& $484,121$& $-,270,355$& $4695$& --- & ---& $8,253,351$& $-6,876,439$& $93$& --- & ---& $21,337,812$& $-10,835,792$\\
   $1$ & $60$& $58$& $58$& $245,479$\cellcolor[gray]{0.8}& $-,131,010$\cellcolor[gray]{0.8}& $5572$& $2084$& $2084$& $4,525,629$\cellcolor[gray]{0.8}& $-5,834,112$\cellcolor[gray]{0.8}& $93$& $1015$& $1015$& $14,592,018$\cellcolor[gray]{0.8}& $-7,813,217$\cellcolor[gray]{0.8}\\
   $2$ & $62$& $29$& $26$& $366,351$& $-,201,047$& $5431$& $3947$& $3947$& $7,503,859$& $-6,311,244$& $95$& $1094$& $2541$& $14,657,975$& $-8,185,791$\\
   $3$ & $50$& $11$&  $7$& $644,083$& $-,332,889$& $1899$& $829$& $783$& $12,527,730$& $-9,776,214$& $92$& $1225$& $1896$& $16,482,714$& $-8,669,765$\\
  \end{tabular}}

  \caption{\label{tab:dl_net}{Summary of inference results for empirical
  temporal networks.} Description length $\Sigma = -\log_2
  P(\{(i,j)_t\},c,b)$ in bits as well as inferred number of node groups
  $C$, token groups $B_\text{N}$, and memory groups $B_{\text{M}}$ for
  different data sets and different Markov order $n$ (see Methods
  section Datasets). The value $-\Delta\Sigma \leq \ln P(\{x_t'\} |
  \{x_t^*\}, b^*)$ is a lower-bound on the predictive likelihood of the
  validation set $\{x_t'\}$ given the training set $\{x_t^*\}$ and its
  best parameter estimate. Values in grey correspond to the minimum of
  each column.}
\end{table*}

A general model for temporal networks treats the edge
sequence as a time series~\cite{holme_temporal_2012,
holme_modern_2015,scholtes_causality-driven_2014}. We can in principle
use the present model without any modification by considering the observed
edges as tokens in the Markov chain, that is, $x_t=(i,j)_t$, where $i$ and
$j$ are the endpoints of the edge at time $t$ (see
Fig.~\ref{fig:diagram}b). However, this can be suboptimal if the networks are
sparse, that is, if only a relatively small subset of all possible edges
occur, and thus there are insufficient data to reliably fit the
model. Therefore, we adapt the model above by including an additional
generative layer between the Markov chain and the observed edges. We do
so by partitioning the {nodes} of the network into groups,
that is, $c_i\in[1,C]$ determines the membership of node $i$ in one of $C$
groups, such that each edge $(i,j)$ is associated with a label
$(c_i,c_j)$. Then we define a Markov chain for the sequence
of {edge labels} and sample the actual edges conditioned
only on the labels. Since this reduces the number of possible tokens
from $O(N^2)$ to $O(C^2)$, it has a more controllable number of
parameters that can better match the sparsity of the data.  We further
assume that, given the node partitions, the edges themselves are sampled
in a degree-corrected manner, conditioned on the edge labels,
\begin{equation}
  P((i, j) | (r,s), \kappa, c) = \begin{cases}
    \delta_{c_i,r}\delta_{c_j,s}\kappa_i\kappa_j  & \text{ if } r\neq s \\
    2 \delta_{c_i,r}\delta_{c_j,s}\kappa_i\kappa_j  & \text{ if } r=s,
    \end{cases}
\end{equation}
where $\kappa_i$ is the probability of a node being selected inside a
group, with $\sum_{i\in r}\kappa_i = 1$. The total likelihood
conditioned on the label sequence becomes
\begin{equation}
  P(\{(i,j)_t\} | \{(r,s)_t\}, \kappa, c) = \prod_tP((i, j)_t | (r,s)_t, \kappa).
\end{equation}
Since we want to avoid overfitting the model, we once more use
noninformative priors, but this time on $\{\kappa_i\}$, and integrate over
them,
\begin{multline}
  P(\{(i,j)_t\} | \{(r,s)_t\}, c)\\
  = \int P(\{(i,j)_t\} | \{(r,s)_t\}, \kappa, c)P(\kappa)\,\dd\kappa.
\end{multline}
Combining this result with Eq.~\ref{eq:chain_dl}, we have the complete
likelihood of the temporal network,
\begin{equation}
  P(\{(i,j)_t\}|c,b)=P(\{(i,j)_t\}|\{(r,s)_t\},c)P(\{(r,s)_t\}|b),
\end{equation}
conditioned only on the partitions.  As we show in detail in
the Methods section Temporal networks, this model is a direct generalisation of
the static DCSBM, with a likelihood composed of two separate static and
dynamic terms. One recovers the static DCSBM exactly by choosing
$B_\text{N}=B_{\text{M}}=1$---making the state transitions memoryless.

Finally, to make the model nonparametric, we again include the same
prior as before for the node partition $c$, in addition to token and memory
partition $b$, such that the total nonparametric joint likelihood is
maximised,
\begin{equation}
  P(\{(i,j)_t\},c,b)=P(\{(i,j)_t\}|c,b)P(c)P(b).
\end{equation}
In this way, we again protect against overfitting, and we can infer
not only the number of memory groups $B_\text{N}$ and token groups $B_{\text{M}}$, as
before, but also the number of groups in the temporal network itself,
$C$. If, for example, the temporal network is completely random---that is,
the edges are placed randomly both in the aggregated network as
well as in time---we again infer $B_\text{N}=B_{\text{M}}=C=1$ with the largest
probability. We refer to the Methods section Temporal networks for a complete
derivation of the final likelihood.

\begin{figure*}
  \begin{overpic}[width=.9\textwidth]{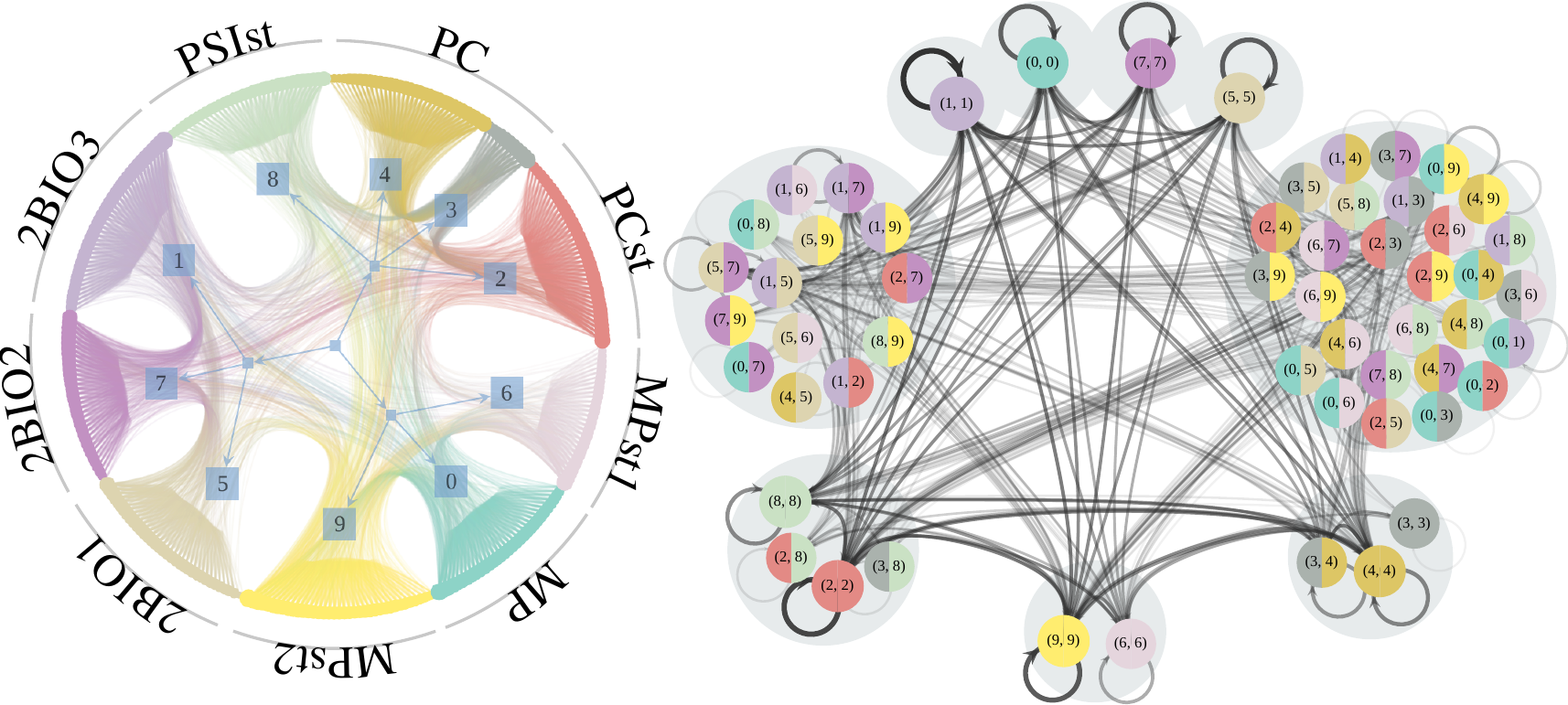}
    \put(0,40){\larger\boldlabel{a}}
    \put(45,40){\larger\boldlabel{b}}
  \end{overpic}
  \caption{\label{fig:thiers}{
  Inferred temporal model for a high school proximity network~\cite{mastrandrea_contact_2015}}. (\boldlabel{a}) The static part of the
  model divides the high school students into $C=10$ groups (square nodes) that
  almost match the known classes (text labels).
  (\boldlabel{b}) The dynamic part of the model divides the directed multigraph
  group pairs in \boldlabel{a} into $B_\text{N}=B_{\text{M}}=9$ groups (grey circles).  The model
  corresponds to an $n=1$ unified Markov chain on the edge labels, where
  the memory and tokens have identical partitions, as described in detail in the Methods section The unified first-order model.}
\end{figure*}

We employ this model in a variety of dynamic network datasets from
different domains (for details, see Table~\ref{tab:dl_net} and
Methods section Datasets). In all cases, we
infer models with $n>0$ that identify many groups for the tokens and
memories, meaning that the model succeeds in capturing temporal
structures. In most cases, models with $n=1$ best describe the data,
implying that there is not sufficient evidence for higher-order memory,
with exception of the network of chess moves, which is best
described by a model with $n=2$. This result is different from the
results for the comparably long non-network sequences in
table~\ref{tab:dl_seq}, where we identified higher-order Markov
chains. Again, this is because the alphabet size is much larger for
temporal networks---corresponding to all possible edges that can be
encountered. Hence, for the datasets in Table~\ref{tab:dl_net} the size
of the alphabet is often comparable with the length of the sequence. In
view of this, it is remarkable that the method can detect any structure
at all. The intermediary layer where the Markov chain generates edge
types instead of the edges directly is crucial.  If we fit the original
model without this modification, we indeed get much larger description
lengths and we often fail to detect any Markov structure (not shown).

To illustrate how the model characterises the temporal structure of
these systems, we focus on the proximity network of high school
students, which corresponds to the voluntary tracking of $327$ students
for a period of $5$ days~\cite{mastrandrea_contact_2015}. Whenever the
distance between two students fell below a threshold, an edge between
them was recorded at that time. In the best-fitting model for these data,
the inferred groups for the aggregated
network correspond exactly to the known division into 9 classes,
except for the PC class, which was divided into two groups (Fig.~\ref{fig:thiers}).
The groups show a clear assortative structure,
where most connections occur within each class. The clustering of the
edge labels in the second part of the model reveals the temporal
dynamics. We observe that the edges connecting nodes of the same group
cluster either in single-node or small groups, with a high incidence of
self-loops. This means that if an edge that connects two students of the
same class appears in the sequence, the next edge is most likely also
inside the same class, indicating that the students of the same class
are clustered in space and time. The remaining edges between students of
different classes are separated into two large groups. This division
indicates that the different classes meet each other at different
times. Indeed, the classes are located in different parts of the school
building and they typically go to lunch separately~\cite{mastrandrea_contact_2015}.
Accordingly, our method can uncover the associated dynamical pattern from the data alone.

\subsection*{Temporal prediction}

Using generative models to extract patterns in data yields more
than a mere description, since they generalise 
observations and predict future events. For our
Bayesian approach, the models can even be used to predict tokens and
memories not previously observed. This ability is in strong contrast to more
heuristic methods that are only designed to find partitions in networks
or time series, and cannot be used for prediction. Furthermore,
the predictive performance of a model is often used on its own to
evaluate it, and serves as an alternative approach to model selection:
since an overly complicated model incorporates noise in its description,
it yields less accurate predictions. Thus, maximizing the predictive
performance also amounts to a balance between quality of fit and model
complexity, similarly to the minimum description length approach we have
used so far. It is important, therefore, to determine if these two
different criteria yield consistent results, which would serve as an
additional verification of the overall approach.

We show this consistency by considering a scenario where a
sequence is divided into two equal-sized contiguous parts, $\{x_t\} =
\{x^*_t\} \cup \{x_t'\}$. That is, a training set $\{x^*_t\}$ and a
validation set $\{x_t'\}$. We then evaluate the model by fitting it to
the training set and use it to predict the validation set.  If we
observe only the training set, the likelihood of the
validation set is bounded below by $P(\{x_t'\} | \{x_t^*\}, b^*)
\ge \exp(-\Delta\Sigma)$, where $b^* =\operatorname{argmax}_b P(b | \{x_t^*\})$
is the best partition given the training set and $\Delta\Sigma$ is
the description length difference between the training set and
the entire data (for a proof, see Methods section Predictive held-out likelihood). This lower
bound will become tight when both the validation and training sets
become large, and hence can be used as an asymptotic approximation of
the predictive likelihood. Table~\ref{tab:dl_net} shows empirical values
for the same datasets as considered before, where $n=0$ corresponds to
using only the static DCSBM to predict the edges, ignoring any time
structure. The temporal network model provides better prediction in all
cases, and the best Markov order always coincides with the one that
yields the minimum description length, thus confirming a full agreement
between both criteria in these cases.

\section*{Discussion}\label{sec:conclusion}

We presented a dynamical variation of the degree-corrected stochastic
block model that can capture long pathways or large-scale structures
in sequences and temporal networks. The model does not require the
optimal Markov order or number of groups as inputs, but infers them
from data because the underlying arbitrary-order Markov chain model is
nonparametric. Its nonparametric nature also evades a priori imposed
timescales. We showed that the model successfully finds meaningful
large-scale temporal structures in real-world systems and that it
predicts their temporal evolution. Moreover, in the Methods section 
we extend the model to situations where the dynamics take place in
continuous time or is nonstationary.
In contrast to approaches that force network-formation
dynamics into discrete time windows, and require a priori knowledge
about the appropriate amount of dynamical memory, our approach
provides a principled and versatile alternative.

\bibliographystyle{naturemag_noURL}
\bibliography{bib}

\section*{Methods}

\subsection*{Bayesian Markov chains with communities}\label{sec:communities}

As described in the main text, a Bayesian formulation of the Markov
model consists in specifying prior probabilities for the model
parameters, and integrating over them. In doing so, we convert the
problem from one of parametric inference where the model
parameters need to be specified before inference, to a nonparametric
one where no parameters need to be specified before inference. In this
way, the approach possesses intrinsic regularisation, where the order of
the model can be inferred from data alone, without
overfitting~\cite{jaynes_probability_2003, mackay_information_2003}.

To accomplish this, we rewrite the model likelihood, using
Eqs.~\ref{eq:Markov} and~\ref{eq:block_prob}, as
\begin{equation}
  P(\{x_t\} |b, \lambda,\theta) = \prod_{x,\bm{x}} \left(\theta_x\lambda_{b_x,b_{\bm{x}}}\right)^{a_{x,\bm{x}}} = \prod_{x}\theta_x^{k_x} \prod_{r<s}\lambda_{rs}^{e_{rs}},
\end{equation}
and observe the normalisation constraints $\sum_{x\in r}\theta_x = 1$,
and $\sum_r\lambda_{rs} = 1$. Since this is just a product of
multinomials, we can choose conjugate Dirichlet priors probability
densities $\mathcal{D}_r(\{\theta_x\}|\{\alpha_x\})$ and
$\mathcal{D}_s(\{\lambda_{rs}\}|\{\beta_{rs}\})$, which allows us to exactly compute the integrated likelihood,
\begin{multline}\label{eq:int_likelihood}
  P(\{x_t\}|\alpha,\beta,b) = \int \dd\theta \dd\lambda\; P(\{x_t\} |b,\lambda,\theta)  \\ \times \prod_r\mathcal{D}_r(\{\theta_x\}|\{\alpha_x\}) \prod_s \mathcal{D}_s(\{\lambda_{rs}\}|\{\beta_{rs}\}) \\
  = \left[\prod_r \frac{\Gamma(A_r)}{\Gamma(e_r + A_r)}\prod_{x\in r}\frac{\Gamma(k_x + \alpha_x)}{\Gamma(\alpha_x)}\right] \\  \times \left[\prod_s\frac{\Gamma(B_s)}{\Gamma(e_s + B_s)}\prod_r\frac{\Gamma(e_{rs} + \beta_{rs})}{\Gamma(\beta_{rs})}\right],
\end{multline}
where $A_r = \sum_{x\in r}\alpha_x$ and $B_s = \sum_r\beta_{rs}$.  We
recover the Bayesian version of the common Markov chain formulation (see
Ref.~\cite{strelioff_inferring_2007}) if we put each memory and token in
their own groups. This remains a parametric distribution, since we need
to specify the hyperparameters. However, in the absence of prior
information it is more appropriate to make a noninformative choice that
encodes our a priori lack of knowledge or preference towards any
particular model, which amounts to choosing $\alpha_x=\beta_{rs}=1$,
making the prior distributions flat. If we substitute these values in
Eq.~\ref{eq:int_likelihood}, and re-arrange the terms, we can show that
it can be written as the following combination of conditional
likelihoods,
\begin{multline}
  P(\{x_t\}|b, \{e_s\}) =  P(\{x_t\}|b,\{e_{rs}\}, \{k_x\}) \\
  \times P(\{k_x\}|\{e_{rs}\},b)P(\{e_{rs}\}|\{e_s\}),
\end{multline}
where
\begin{align}
  P(\{x_t\}|b,\{e_{rs}\}, \{k_x\}) &= \frac{\prod_{r<s} e_{rs}!}{\prod_r e_r!\prod_s e_s!}\prod_xk_x!, \label{eq:micro_l}\\
  P(\{k_x\}|\{e_{rs}\}, b) &= \left[\prod_r\multiset{n_r}{e_r}\right]^{-1},\label{eq:prior_pk}\\
  P(\{e_{rs}\}|\{e_s\}) &= \left[\prod_s\multiset{B_\text{N}}{e_s}\right]^{-1},\label{eq:prior_ers}
\end{align}
with $\multiset{m}{n}={m+n-1\choose n}$ being the multiset coefficient,
that counts the number of $m$-combinations with repetitions from a set
of size $n$.  The expression above has the following combinatorial
interpretation: $P(\{x_t\}|b,\{e_{rs}\}, \{k_x\})$ corresponds to the
likelihood of a {microcanonical}
model~\cite{peixoto_nonparametric_2017} where a random sequence
$\{x_t\}$ is produced with exactly $e_{rs}$ total transitions between
groups $r$ and $s$, and with each token $x$ occurring exactly $k_x$
times. In order to see this, consider a chain where there are only $e_{rs}$
transitions in total between token group $r$ and memory group $s$, and
each token $x$ occurs exactly $k_x$ times. For the first transition
in the chain, from a memory $\bm{x}_0$ in group $s$ to a token $x_1$ in
group $r$, we have the probability
\begin{equation}
  P(x_1|\bm{x}_0, b, \{e_{rs}\}, \{k_x\}) = \frac{e_{rs} k_{x_1}}{e_s e_r}.
\end{equation}
Now, for the second transition from memory $\bm{x}_1$ in group
$t$ to a token $x_2$ in group $u$, we have the probability
\begin{multline}
  P(x_2|\bm{x}_1, b, \{e_{rs}\}, \{k_x\}) = \\ \qquad
  \begin{cases}\displaystyle
    \frac{e_{ut} k_{x_2}}{e_t e_u}, & \text{ if } t \neq s ,\; u \neq r ,\; x_2 \neq x_1, \\ \displaystyle
    \frac{(e_{us} - 1) k_{x_2}}{(e_s-1) e_u}, &\text{ if } t = s ,\; u \neq r ,\; x_2 \neq x_1, \\ \displaystyle
    \frac{e_{rt} (k_{x_1} - 1)}{e_t (e_r-1)}, &\text{ if } t \neq s ,\; u = r ,\; x_2 = x_1, \\ \displaystyle
    \frac{e_{rt} k_{x_2}}{e_t (e_r-1)}, &\text{ if } t \neq s ,\; u = r ,\; x_2 \neq x_1, \\ \displaystyle
    \frac{(e_{rs}-1)k_{x_2}}{(e_s-1)(e_r-1)}, &\text{ if } t = s ,\; u = r ,\; x_2 \neq x_1, \\ \displaystyle
    \frac{(e_{rs}-1) (k_{x_1}-1)}{(e_s - 1) (e_r-1)}, &\text{ if } t = s ,\; u = r ,\;   x_2 = x_1.\\
  \end{cases}
\end{multline}
Proceeding recursively, the final likelihood for the entire chain is
\begin{equation}
  P(\{x_t\}|b,\{e_{rs}\}, \{k_x\}) = \frac{\prod_{rs} e_{rs}!}{\prod_r e_r!\prod_s e_s!}\prod_xk_x!,
\end{equation}
which is identical to Eq.~\ref{eq:micro_l}.

The remaining terms in Eqs.~\ref{eq:prior_pk} and~\ref{eq:prior_ers} are
the prior probabilities on the discrete parameters $\{k_x\}$ and
$\{e_{rs}\}$, respectively, which are uniform distributions of the type
$1/\Omega$, where $\Omega$ is the total number of possibilities given
the imposed constraints. We refer to
Ref.~\cite{peixoto_nonparametric_2017} for a more detailed discussion on
those priors.

Since the integrated likelihood above gives $P(\{x_t\}|b, \{e_s\})$, we
still need to include priors for the node partitions $\{b_x\}$ and
$\{b_{\bm{x}}\}$, as well as memory group counts, $\{e_s\}$, to make
the above model fully nonparametric. This is exactly the same situation
encountered with the
SBM~\cite{peixoto_parsimonious_2013,peixoto_hierarchical_2014,peixoto_nonparametric_2017}.
Following
Refs.~\cite{peixoto_hierarchical_2014,peixoto_nonparametric_2017}, we
use a nonparametric two-level Bayesian hierarchy for the partitions,
$P(\{b_i\}) = P(\{b_i\}|\{n_r\})P(\{n_r\})$, with uniform distributions
\begin{equation}\label{eq:part}
  P(\{b_i\}|\{n_r\}) = \frac{\prod_rn_r!}{M!},\quad P(\{n_r\}) = {M-1\choose B-1}^{-1},
\end{equation}
where $n_r$ is the number of nodes in group $r$, $M=\sum_rn_r$, which we
use for both $\{b_x\}$ and $\{b_{\bm{x}}\}$, that is, $P(b) =
P(\{b_x\})P(\{b_{\bm{x}}\})$. Analogously, for $\{e_s\}$ we can use a
uniform distribution
\begin{equation}
  P(\{e_s\}|b) = \multiset{B_{\text{M}}}{E}^{-1}.
\end{equation}
The above priors make the model fully nonparametric with a
joint and marginal probability $P(\{x_t\},b) = P(\{x_t\},b,\{e_s\}) =
P(\{x_t\}|b,\{e_s\})P(b)P(\{e_s\})$. 
This approach successfully finds the most appropriate number of groups according to statistical evidence, without
overfitting~\cite{peixoto_parsimonious_2013,peixoto_hierarchical_2014,peixoto_model_2015,rosvall_information-theoretic_2007}. This
nonparametric method can also detect the most appropriate order of
the Markov chain, again without overfitting~\cite{strelioff_inferring_2007}.  However, in some ways it
is still sub-optimal. The use of conjugate Dirichlet priors above was
primarily for mathematical convenience, not because they closely
represent the actual mechanisms believed to generate the data. Although
the noninformative choice of the Dirichlet distribution (which yields
flat priors for $\{e_{rs}\}$ and $\{e_s\}$) can be well
justified by maximum entropy arguments (see
Ref.~\cite{jaynes_probability_2003}), and are unbiased, it can in fact
be shown that it can lead to {underfitting} of the data, where the
maximum number of detectable groups scales sub-optimally as
$\sqrt{N}$~\cite{peixoto_parsimonious_2013}. As shown in
Ref.~\cite{peixoto_hierarchical_2014}, this limitation can be overcome
by departing from the model with Dirichlet priors, and replacing
directly the priors $P(\{e_{rs}\}|\{e_s\})$ and $P(\{e_s\})$ of the
microcanonical model by a single prior $P(\{e_{rs}\})$, and noticing
that $\{e_{rs}\}$ corresponds to the adjacency matrix of bipartite
multigraph with $E$ edges and $B_\text{N}+B_{\text{M}}$ nodes.  With this insight, we
can write $P(\{e_{rs}\})$ as a Bayesian hierarchy of nested SBMs, which
replaces the resolution limit above by $N/\ln N$, and provides a
multilevel description of the data, while remaining unbiased.
Furthermore, the uniform prior in Eq.~8 for the token frequencies
$P(\{k_x\}|\{e_{rs}\}, b)$ intrinsically favours concentrated
distributions of $k_x$ values. 
This distribution is often skewed. We therefore replace it by a
two-level Bayesian hierarchy $P(\{k_x\}|\{e_{rs}\}, b) = \prod_r
P(\{k_x\}|\{n_k^r\})P(\{n_k^r\}|e_r)$, with
\begin{equation}
  P(\{k_x\}|\{n_k^r\}) = \frac{\prod_k n^r_k!}{n_r!},
\end{equation}
and $P(\{n_k^r\}|e_r) = q(e_r, n_r)^{-1}$, where $q(m,n)$ is the number
of restricted partitions of integer $m$ into at most $n$ parts (see
Ref.~\cite{peixoto_nonparametric_2017} for details).

As mentioned in the main text, in order to fit the model above we need
to find the partitions $\{b_x\}$ and $\{b_{\bm{x}}\}$ that maximise
$P(\{x_t\}, b)$, or fully equivalently, minimise the description length
$\Sigma = -\ln P(\{x_t\}, b)$~\cite{grunwald_minimum_2007}. Since this
is functionally equivalent to inferring the DCSBM in networks, we can
use the same algorithms. In this work we employed the fast multilevel
MCMC method of Ref.~\cite{peixoto_efficient_2014}, which has log-linear
complexity $O(N\log^2 N)$, where $N$ is the number of nodes (in our
case, memories and tokens), independent of the number of groups.

\subsection*{Code availability}\label{sec:code}

A free C++ implementation of the inference algorithm is available as
part of the \texttt{graph-tool} Python
library~\cite{peixoto_graph-tool_2014}, available at
\url{http://graph-tool.skewed.de}.

\subsection*{The unified first-order model}\label{sec:unified}

The model defined in the main text is based on a co-clustering of memory
and tokens. In the $n=1$ case, each memory corresponds to a single
token. In this situation, we consider a slight variation of the model
where we force the number of groups of each type to be the same,
that is, $B_\text{N}=B_{\text{M}}=B$, and both partitions to be identical. Instead of
clustering the original bipartite graph, this is analogous to clustering
its projection into a directed transition graph with each node
representing a specific memory and token simultaneously.  When
considering this model, the likelihoods computed in the main text and
above remain exactly the same, with the only difference that we
implicitly force both memory and token partitions to be identical, and
omit the partition likelihood of Eq.~\ref{eq:part} for one of them. We
find that for many datasets this variation provides a slightly better
description than the co-clustering version, although there are also
exceptions to this.

We used this variation of the model in Fig.~\ref{fig:thiers} because it
yielded a smaller description length for that dataset, and simplified
the visualisation and interpretation of the results in that particular
case.

\subsection*{Temporal networks}\label{sec:temporal_app}

Here we show in more detail how the likelihood for the temporal network
model is obtained. As we discuss in the Results section Temporal networks, the total
likelihood of the network conditioned on the label sequence
is
\begin{multline}
  P(\{(i,j)_t\} | \{(r,s)_t\}, \kappa, c) = \prod_tP((i, j)_t | (r,s)_t, \kappa) \\
  = \left[\prod_t\delta_{c_{i_t},r_t}\delta_{c_{j_t},s_t} \right] \prod_i \kappa_i^{d_i} \prod_r2^{m_{rr}},
\end{multline}
where $d_i$ is the degree of node $i$, and $m_{rs}$ is the total number
of edges between groups $r$ and $s$.  Maximum likelihood estimation gives $\hat\kappa_i = d_i/e_{c_i}$. But since we want to avoid
overfitting the model, we once more use noninformative priors, this
time on $\{\kappa_i\}$, integrate over them, 
henceforth omitting the trivial Kronecker delta term above and obtain
\begin{equation}
  P(\{(i,j)_t\} | \{(r,s)_t\}, c) = \frac{\prod_id_i!\prod_r2^{m_{rr}}}{\prod_re_r!}P(\{d_i\}),
\end{equation}
with $P(\{d_i\})=\prod_r\multiset{n_r}{e_r}^{-1}$.  Combining this with
Eq.~\ref{eq:chain_dl} as
$P(\{(i,j)_t\}|c,b)=P(\{(i,j)_t\}|\{(r,s)_t\},c)P(\{(r,s)_t\}|b)$, we
have the complete likelihood of the temporal network
\begin{align}
  P(\{(i,j)_t\}|c,b)= \frac{\prod_{r\ge s}m_{rs}!\prod_r2^{m_{rr}}}{\prod_re_r!}\prod_id_i! \\
  \times P(\{d_i\}|c)P(\{m_{rs}\}) \frac{\prod_{u<v}e'_{uv}!}{\prod_ue'_u!\prod_ve'_v!}P(\{e'_{uv}\}).
\end{align}
This likelihood can be rewritten in such a way that makes clear that it
is composed of one purely static and one purely dynamic part,
\begin{multline}\label{eq:dynamic_net_dl}
  P(\{(i,j)_t\}|c,b) = P(\{A_{ij}\}|c) \times \frac{P(\{(r,s)_t\}|b,\{e_v\})}{P(\{m_{rs}\})\prod_{r\ge s}m_{rs}!}.
\end{multline}
The first term of Eq.~\ref{eq:dynamic_net_dl} is precisely the
nonparametric likelihood of the {static} DCSBM that generates the
{aggregated graph} with adjacency matrix $A_{ij} = k_{x=(i,j)}$
given the node partition $\{c_i\}$, which itself is given by
\begin{multline}
  \ln P(\{A_{ij}\}|c) \approx E + \frac{1}{2}\sum_{rs}e_{rs}\ln\frac{e_{rs}}{e_re_s} + \sum_i\ln d_i! \\
  + \ln P(\{d_i\}) + \ln P(\{m_{rs}\}),
\end{multline}
if Stirling's approximation is used. The second term in
Eq.~\ref{eq:dynamic_net_dl} is the likelihood of the Markov chain of
edge labels given by Eq.~\ref{eq:chain_dl} (with $\{x_t\}=\{(r,s)_t\}$,
and $\{k_x\} = \{m_{rs}\}$).  This model, therefore, is a direct
generalisation of the static DCSBM, with a likelihood composed of two
separate static and dynamic terms. One recovers the static DCSBM exactly
by choosing $B_\text{N}=B_{\text{M}}=1$---making the state transitions memoryless---so
that the second term in Eq.~\ref{eq:dynamic_net_dl} above contributes
only with a trivial constant $1/E!$ to the overall
likelihood. Equivalently, we can view the DCSBM as a special case with
$n=0$ of this temporal network model.

\subsection*{Predictive held-out likelihood}\label{sec:held-out}

Given a sequence divided in two contiguous parts, $\{x_t\} = \{x^*_t\}
\cup \{x_t'\}$, that is, a training set $\{x^*_t\}$ and a validation set
$\{x_t'\}$, and if we observe only the training set, the predictive
likelihood of the validation set is
\begin{equation}
  P(\{x_t'\} | \{x_t^*\}, b^*) = \frac{P(\{x_t'\} \cup \{x_t^*\} | b^*)}{P(\{x_t^*\} | b^*)},
\end{equation}
where $b^* =\operatorname{argmax}_b P(b | \{x_t^*\})$
is the best partition given the training set. Moreover, we have
\begin{equation}
  P(\{x_t'\} \cup \{x_t^*\} | b^*) = \sum_{b'}  P(\{x_t'\} \cup \{x_t^*\} | b^*, b') P(b'|b^*),
\end{equation}
where $b'$ corresponds to the partition of the newly observed memories
(or even tokens) in $\{x_t'\}$. Generally we have $P(b'|b^*) = P(b',
b^*)/P(b^*)$, so that
\begin{multline}
  P(\{x_t'\} | \{x_t^*\}, b^*) = \frac{\sum_{b'}  P(\{x_t'\} \cup \{x_t^*\} | b^*, b') P(b^*, b')}{P(\{x_t^*\} | b^*)P(b^*)}\\
   \ge \frac{P(\{x_t'\} \cup \{x_t^*\} | b^*, \hat b') P(b^*, \hat b')}{P(\{x_t^*\} | b^*)P(b^*)} = \exp(-\Delta\Sigma),
\end{multline}
where $ \hat b' = \operatorname{argmax}_{b'} P(\{x_t'\} \cup \{x_t^*\} |
b^*, b') P(b^*, b')$ and $\Delta\Sigma$ is the difference in the
description length between the training set and the entire data. Hence,
computing the minimum description length of the remaining data by
{maximising} the posterior likelihood relative to the
partition of the previously unobserved memories or tokens, yields a
{lower bound} on the predictive likelihood. This lower bound will
become tight when both the validation and training sets become large, because
then the posterior distributions concentrate around the maximum, and hence can be used as an asymptotic approximation of the predictive likelihood.

\subsection*{Continuous time}
So far, we have considered sequences and temporal networks that evolve
discretely in time. Although this is the appropriate description for
many types of data, such as text, flight itineraries and chess moves, in
many other cases events happen instead in real time. In this case, the
time series can be represented---without any loss of generality---by
an embedded sequence of tokens $\{x_t\}$ placed in discrete time,
together with an additional sequence of {waiting times}
$\{\Delta_t\}$, where $\Delta_t \ge 0$ is the real time difference
between tokens $x_t$ and $x_{t-1}$. Employing a continuous-time Markov
chain description, the data likelihood can be written as
\begin{equation}
  P(\{x_t\},\{\Delta_t\}|p,\lambda)=P(\{x_t\}|p)\times P(\{\Delta_t\}|\{x_t\},\lambda)
\end{equation}
with $P(\{x_t\}|p)$ given by Eq.~\ref{eq:Markov}, and
\begin{equation}\label{eq:wait}
  P(\{\Delta_t\}|\{x_t\},\lambda) = \prod_tP(\Delta_t|\lambda_{\bm{x}_{t-1}}),
\end{equation}
where
\begin{equation}\label{eq:dt_exp}
P(\Delta|\lambda) = \lambda e^{-\lambda\Delta},
\end{equation}
is a maximum-entropy distribution governing the waiting times, according
to the frequency $\lambda$. Substituting this in Eq.~\ref{eq:wait}, we
have
\begin{equation}
  P(\{\Delta_t\}|\{x_t\},\lambda) = \prod_{\bm{x}}\lambda_{\bm{x}}^{k_{\bm{x}}}e^{-\lambda_{\bm{x}}\Delta_{\bm{x}}},
\end{equation}
where $\Delta_{\bm{x}} = \sum_t\Delta_t\delta_{\bm{x}_t,\bm{x}}$. To
compute the nonparametric Bayesian evidence, we need a
conjugate prior for the frequencies $\lambda_{\bm{x}}$,
\begin{equation}\label{eq:lambda_prior}
  P(\lambda|\alpha,\beta) = \frac{\beta^\alpha\lambda^{\alpha-1}}{\Gamma(\alpha)}e^{-\beta\lambda},
\end{equation}
where $\alpha$ and $\beta$ are hyperparameters, interpreted,
respectively, as the number and sum of prior observations. A fully
noninformative choice would entail $\alpha\to 0$ and $\beta\to 0$, which
would yield the so-called Jeffreys prior~\cite{jeffreys_invariant_1946},
$P(\lambda) \propto 1/\lambda$. Unfortunately, this prior is improper because it is not a normalised distribution. In order to avoid this, we use
instead $\alpha = 1$ and $\beta = \sum_{\bm{x}}{\lambda_{\bm{x}}}/M$,
taking into account the global average. While this is not the only
possible choice, the results should not be sensitive to this prior since the data will eventually override any reasonable assumption we make. Using this prior, we obtain the Bayesian
evidence for the waiting times as
\begin{align}
  P(\{\Delta_t\}|\{x_t\}) &= \prod_{\bm{x}}\int_0^{\infty}\lambda^{k_{\bm{x}}}e^{-\lambda\Delta_{\bm{x}}}P(\lambda|\alpha,\beta)\,\dd\lambda,\\
                          &=\prod_{\bm{x}} \frac{\beta^\alpha\Gamma(k_{\bm{x}}+\alpha)}{\Gamma(\alpha)(\Delta_{\bm{x}}+\beta)^{k_{\bm{x}}+\alpha}}.
\end{align}
Hence, if we employ the Bayesian parametrisation with communities for
the discrete embedded model as we did previously, we have
\begin{equation}\label{eq:evidence_wait}
  P(\{x_t\},\{\Delta_t\},b) = P(\{x_t\},b)\times P(\{\Delta_t\}|\{x_t\}),
\end{equation}
with $P(\{x_t\},b)$ given by Eq.~\ref{eq:joint_complete}.

Since the partition of memories and tokens only influences the first
term of~Eq.~\ref{eq:evidence_wait}, corresponding to the embedded
discrete-time Markov chain, $P(\{x_t\},b)$, the outcome of the inference
for any particular Markov order will not take into account the
distribution of waiting times---although the preferred Markov order
might be influenced by it. We can change this by modifying the model
above, assuming that the waiting times are conditioned on the group
membership of the memories,
\begin{equation}
  \lambda_{\bm{x}} = \eta_{b_{\bm{x}}},
\end{equation}
where $\eta_r$ is a frequency associated with memory group $r$. The
Bayesian evidence is computed in the same manner, integrating over
$\eta_r$ with the noninformative prior of Eq.~\ref{eq:lambda_prior},
yielding
\begin{equation}
  P(\{\Delta_t\}|\{x_t\}) = \prod_r \frac{\beta^\alpha\Gamma(e_r+\alpha)}{\Gamma(\alpha)(\Delta_r+\beta)^{e_r+\alpha}},
\end{equation}
where $\Delta_r = \sum_t\Delta_t\delta_{b_{\bm{x}_t},r}$. Since this
assumes that the waiting times will be sampled from the same
distribution inside each group, the inference procedure will take the
waiting time into account, and will place memories with significantly different
delays into different groups.

\begin{figure}
  \begin{tabular}{r}
    \begin{overpic}[width=\columnwidth]{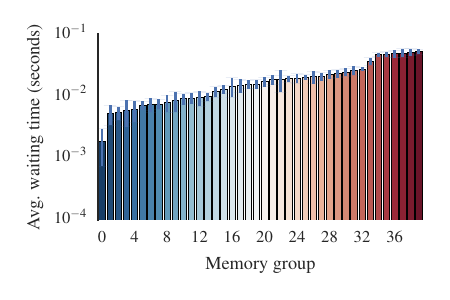}
      \put(0,65){\larger\boldlabel{a}}
    \end{overpic}\\
    \begin{overpic}[width=\columnwidth]{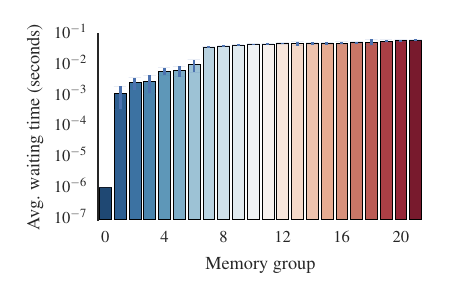}
      \put(0,65){\larger\boldlabel{b}}
    \end{overpic}
  \end{tabular} \caption{Waiting times for a discrete- and a
    continuous-time Markov model. Results inferred
    from Beethoven's fifth symphony. (\boldlabel{a}) $n=1$ discrete-time model,
    ignoring the waiting times between notes. (\boldlabel{b}) $n=2$ continuous-time
    model, with waiting times incorporated into the inference. The error
    bars correspond to the standard deviation of the
    mean.\label{fig:wait}}
\end{figure}

As an example of the use of this model variation, we consider a piano
reduction of Beethoven's fifth symphony (extracted in MIDI format from
the Mutopia project at \url{http://www.mutopiaproject.org}), represented
as a sequence of $E=4,223$ notes of an alphabet of size $N=63$. We
consider both model variants where the timings between notes are
discarded, and where they are included. If individual notes are played
simultaneously as part of a chord, we order them lexicographically and
separate them by $\Delta t = 10^{-6}$ seconds. The results of the
inference can be seen in Table~\ref{tab:dl_cont}. The discrete-time
model favours an $n=1$ Markov chain, whereas the continuous-time model
favours $n=2$. This is an interesting result that shows that the timings
alone can influence the most appropriate Markov order. We can see in
more detail why by inspecting the typical waiting times conditioned on
the memory groups, as shown in Fig.~\ref{fig:wait}. For the
discrete-time model, the actual continuous waiting times (which are not
used during inference) are only weakly correlated with the memory
groups. On the other hand, for the continuous-time model we find that
the memories are divided in such a way that they are strongly correlated
with the waiting times: There is a group of memories for which the
ensuing waiting times are always $\Delta t = 10^{-6}$ --- corresponding
to node combinations that are always associated with chords. The
remaining memories are divided into further groups that display at least
two distinct timescales, that is, short and long pauses between
notes. These statistically significant patterns are only visible for the
higher order $n=2$ model.

\begin{table}
  \begin{tabular}{r|rrc|rrc}\smaller
    &\multicolumn{3}{c|}{Discrete time} & \multicolumn{3}{c|}{Continuous time} \\ \hline \hline
    $n$
    & $B_\text{N}$ & $B_{\text{M}}$ & $-\ln P(\{x_t\},b)$ &  $B_\text{N}$ & $B_{\text{M}}$ & $-\ln P(\{x_t\},\{\Delta_t\},b)$
    \\ \hline
    $1$ & $40$ & $40$ & $13,736$ \cellcolor[gray]{0.8}&  $37$ & $37$ & $58,128$ \\
    $2$ & $35$ & $34$ & $15,768$ &  $24$ & $22$ & $47,408$ \cellcolor[gray]{0.8}\\
    $3$ & $34$ & $33$ & $24,877$ &  $16$ & $15$ & $54,937$ \\
  \end{tabular}

  \caption{Joint likelihoods for discrete- and
    continuous-time Markov models. Results inferred from
    Beethoven's fifth symphony for different
    Markov orders $n$. Values in grey correspond to the maximum
    likelihood of each column.\label{tab:dl_cont}}
\end{table}

In the above model the waiting times are distributed according to the
exponential distribution of Eq.~\ref{eq:dt_exp}, which has a typical
timescale given by $1/\lambda$. However, one often encounters processes
where the dynamics are {bursty}, that is, the waiting times between
events lack any characteristic scale, and are thus distributed according
to a power-law
\begin{equation}\label{eq:dt_pareto}
  P(\Delta|\beta) = \frac{\beta \Delta_m^\beta}{\Delta^{\beta+1}},
\end{equation}
for $\Delta > \Delta_m$, otherwise $P(\Delta|\beta)=0$. One could in
principle repeat the above calculations with the above distribution to
obtain the inference procedure for this alternative model. However, this
is in fact not necessary, since by making the transformation of variables
\begin{equation}\label{eq:mu}
  \mu = \ln \frac{\Delta}{\Delta_m},
\end{equation}
we obtain for Eq.~\ref{eq:dt_pareto}
\begin{equation}
  P(\mu|\beta) = \beta e^{-\beta\mu},
\end{equation}
which is the same exponential distribution of
Eq.~\ref{eq:dt_exp}. Hence, we need only to perform the transformation
of Eq.~\ref{eq:mu} for the waiting times prior to inference, to use the
bursty model variant, while maintaining the exact same algorithm.

\subsection*{Nonstationarity and hidden contexts}

\begin{figure}
    \resizebox{\columnwidth}{!}{
      \begin{tabular}{rr}
        \begin{overpic}[width=.49\columnwidth]{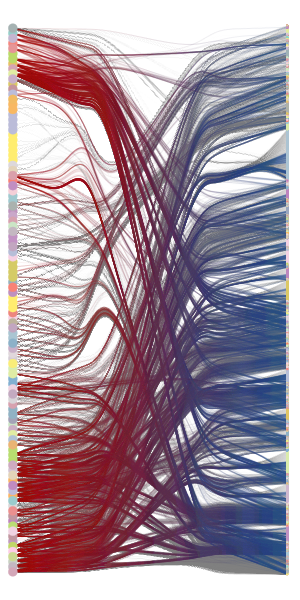}
          \put(1,97){Tokens}
          \put(32.5,97){Memories}
          \put(22,0){\boldlabel{a}}
        \end{overpic}
          &
        \begin{overpic}[width=.49\columnwidth]{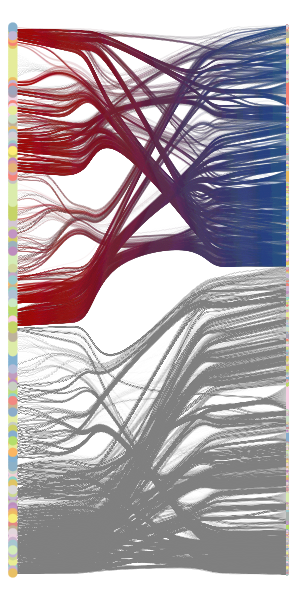}
          \put(1,97){Tokens}
          \put(32.5,97){Memories}
          \put(22,0){\boldlabel{b}}
        \end{overpic}
    \end{tabular}}
    \caption{Markov model fit for a concatenated text. `The texts are `War and peace'' by Leo
      Tolstoy and ``À la recherche du temps perdu'', by Marcel
      Proust. Edge endpoints in red and blue correspond to token and
      memories, respectively, that involve letters exclusive to
      French. (a) Version of the model with $n=3$ where no distinction
      is made between the same token in the different novels,
      yielding a description length $-\log_2 P(\{x_t\},b) =
      7,450,322$. (b) Version with $n=3$ where each token is annotated
      with its novel, yielding a description length $-\log_2
      P(\{x_t\}, b) = 7,146,465$.
      \label{fig:stationarity}}
\end{figure}
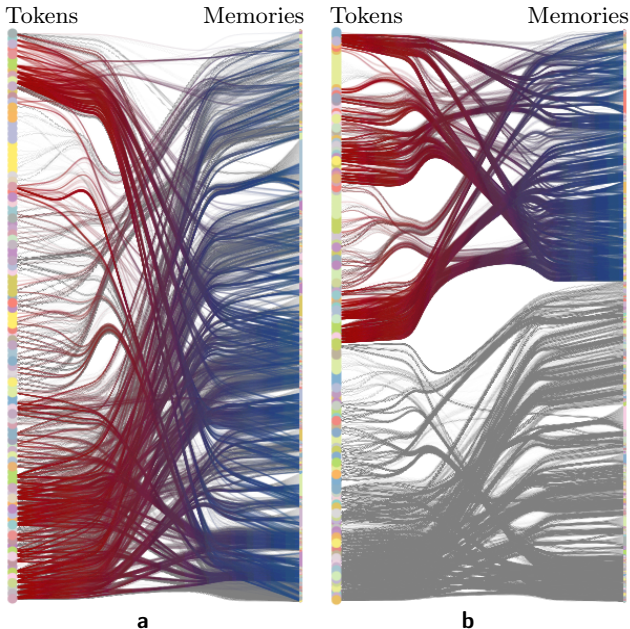

An underlying assumption of the Markov model proposed is that the same
transition probabilities are used for the whole duration of the
sequence, that is, the Markov chain is {stationary}. Generalisations of
the model can be considered where these probabilities change over
time. Perhaps the simplest generalisation is to assume that the dynamics
is divided into $T$ discrete epochs, such that one replaces tokens
$x_t$ by a pair $(x,\tau)_t$, where $\tau\in[1,T]$ represents the epoch
where token $x$ was observed. In fact, $\tau$ does not need to be
associated with a temporal variable---it could be any arbitrary
covariate that describes additional aspects of the data. By
incorporating this type of annotation into the tokens, one can use a
stationary Markov chain describing the augmented tokens that in fact
corresponds to a non-stationary one if one omits the variable $\tau$
from the token descriptors---effectively allowing for arbitrary
extensions of the model by simply incorporating appropriate covariates,
and without requiring any modification to the inference algorithm.

Another consequence of this extension is that the same token $x$ can
belong to different groups if it is associated with two or more
different covariates, $(x,\tau_1)$ and $(x,\tau_2)$. Therefore, this
inherently models a situation where the group membership of tokens and
memory vary in time.

As an illustration of this application of the model, we consider two
literary texts: an English translation of ``War and peace'', by Leo
Tolstoy, and the French original of ``À la recherche du temps perdu'',
by Marcel Proust. First, we concatenate both novels together, treating
it as a single text. If we fit our Markov model to it, we obtain the
$n=3$ model shown in Fig.~\ref{fig:stationarity}a. In that figure, we have
highlighted tokens and memories that involve letters that are exclusive
to the French language, and thus most of them belong to the second
novel. We observe that the model essentially finds a mixture between
English and French. If, however, we indicate in each token to which
novel it belongs, for example $(x, \text{wp})_t$ and $(x, \text{temps})_t$, we
obtain the model of Fig.~\ref{fig:stationarity}b. In this case, the model
is forced to separate between the two novels, and one indeed learns the
French patterns differently from English. Since this nonstationary
model possesses a larger number of memory and tokens, one would expect a
larger description length. However, in this cases it has a
{smaller} description length than the mixed alternative, indicating
indeed that both patterns are sufficiently different to warrant a
separate description. Therefore, this approach is capable of uncovering
{change points}~\cite{peel_detecting_2015}, where the rules
governing the dynamics change significantly from one period to another.

The above extension can also be used to uncover other types of hidden
contexts. For example, in a temporal network of student proximity, we
know that pairs of individuals that are far away are unlikely to be
conditionally dependent on each other. If this spatial information is
not available in the data, it may be inferred in same way it was done
for language above. If the information is available, it can be annotated
on the transitions, yielding a multilayer version of the model, similar
to the layered SBM~\cite{peixoto_inferring_2015}.

\subsection*{Equivalence between structure and dynamics}\label{sec:equiv}

The likelihood of Eq.~4 in the main text is almost the same as the
DCSBM~\cite{karrer_stochastic_2011}. The only exceptions are trivial
additive and multiplicative constants, as well as the fact that the
degrees of the memories do not appear in it. These differences, however,
do not alter the position of the maximum with the respect to the node
partition. This allows us to establish an equivalence between inferring
the community structure of networks and modelling the dynamics taking
place on it. Namely, for a random walk on a connected undirected graph,
a transition $i\to j$ is observed with probability $A_{ij}p_i(t)/k_i$,
with $p_i(t)$ being the occupation probability of node $i$ at time
$t$. Thus, after equilibration with $p_i(\infty)=k_i/2E$, the
probability of observing any edge $(i,j)$ is a constant:
$p_i(\infty)/k_i
+ p_j(\infty)/k_j = 1/E$. Hence, the expected edge counts $e_{rs}$
between two groups in the Markov chain will be proportional to the
actual edge counts in the underlying graph given the same node
partition. This means that the likelihood of Eq.~4 in the main text (for
the $n=1$ projected model described above) and of the DCSBM will differ
only in trivial multiplicative and additive constants, such that the
node partition that maximises them will be identical. This is similar to
the equivalence between network modularity and random
walks~\cite{delvenne_stability_2010}, but here the equivalence is
stronger and we are not constrained to purely assortative modules.
However, this equivalence breaks down for directed graphs, higher order
memory with $n>1$ and when model selection chooses the
number of groups.

\subsection*{Comparison with the map equation for network flows with memory}\label{sec:mapcomp}

Both the community-based Markov model introduced here and the map
equation for network flows with memory~\cite{rosvall_memory_2014}
identify communities in higher-order Markov chains based on maximum
compression. However, the two approaches differ from each other in some
central aspects. The approach presented here is based on the Bayesian
formulation of a generative model, whereas the map equation finds a
minimal modular entropy encoding of the observed dynamics projected on a node
partition. Thus, both approaches seek compression, but of different
aspects of the data.

The map equation operates on the internal and external transitions
within and between possibly nested groups of memory states and describes
the transitions between physical nodes [$x_{t}$ is the physical node or
token in memory states of the form
$\bm{x}=(x_{t},x_{t-1},x_{t-2},\ldots)$].  The description length of
these transitions is minimised for the optimal division of the network
into communities. By construction, this approach identifies assortative
modules of memory states with long flow persistence times. Moreover, for
inferring the most appropriate Markov order, this dynamics approach
requires supervised approaches to model selection that uses random
subsets of the data such as bootstrapping or cross
validation~\cite{persson_maps_2016}.

On the other hand, the model presented here yields a nonparametric
log-likelihood for the entire sequence as well as the model parameters,
with its negative value corresponding to a description length for the
entire data, not only its projection into groups. Minimizing
this description length yields the optimal co-clustering of memories and
tokens, and hence assumes no inherent assortativity. Therefore it can
be used also when the underlying Markov chain is disassortative. Moreover, the
description length can also be used for unsupervised model
selection, where the Markov order and number of groups are determined
from the entire data, obviating the need for bootstrapping or cross
validation. Furthermore, the present approach can be used to generate new data
and make predictions based on past observations.

These distinctions mean that the two different approaches can give
different results and that the problem at hand should decide which method to use.

\subsection*{Datasets}\label{sec:datasets}

Below we give a description of the datasets used in this work.

\begin{description}
  \item[US flight itineraries]{This dataset corresponds to a sample of
  flight itineraries in the US during 2011 collected by Bureau of
  Transportation Statistics of the United States Department of
  Transportation (\url{http://www.transtats.bts.gov/}). The dataset
  contains $1,272,696$ itineraries of varied lengths (airport stops). We
  aggregate all itineraries into a single sequence by concatenating the
  individual itineraries with a special separator token that marks the
  end of a single itinerary. There are $464$ airports in the dataset,
  and hence we have an alphabet of $N=465$ tokens, and a single sequence
  with a total length of $83,653,994$ tokens.}

  \item[War and Peace]{This dataset corresponds to the entire text of
  the english translation of the novel War and Peace by Leo Tolstoy,
  made available by the Project Gutenberg (extracted verbatim from
  \url{https://www.gutenberg.org/cache/epub/2600/pg2600.txt}). This
  corresponds to a sequence with an alphabet of size $N=84$ (including
  letters, space, punctuation and special symbols) and a total length of
  $3,226,652$ tokens.}

  \item[Taxi movements]{This dataset contains GPS logs from $25,000$
  taxi pickups in San Francisco, collected by the company Uber
  (retrieved from
  \url{http://www.infochimps.com/datasets/uber-anonymized-gps-logs},
  also available at
  \url{https://github.com/dima42/uber-gps-analysis}). The geographical
  locations were discretised into $416$ hexagonal cells (see
  Ref.~\cite{rosvall_memory_2014} for details), and the taxi rides were
  concatenated together in a single sequence with a special separator
  token indicating the termination of a ride. In total, the sequence has
  an alphabet of $N=417$ and a length of $819,172$ tokens.}

  \item[RockYou password list]{This dataset corresponds to a widely
  distributed list of $32,603,388$ passwords from the RockYou video game
  company (retrieved from
  \url{http://downloads.skullsecurity.org/passwords/rockyou-withcount.txt.bz2}).
  The passwords were concatenated in a single sequence, with a special
  separator token between passwords. This yields a sequence with an
  alphabet of size $N=215$ (letters, numbers and symbols) and a total
  length of $289,836,299$ tokens.}

  \item[High school proximity]{This dataset corresponds to a temporal
  proximity measurement of students in a french high
  school~\cite{mastrandrea_contact_2015}. A total of $N=327$ students
  were voluntarily tracked for a period of five days, generating
  $E=5,818$ proximity events between pairs of students.}

  \item[Enron email]{This dataset corresponds to time-stamped collection
  of $E=1,148,072$ emails between directed pairs of $N=87,273$, senders
  and recipients of the former Enron corporation, disclosed as part of a
  fraud investigation~\cite{klimt_enron_2004}.}

  \item[Internet AS]{This dataset contains connections between
  autonomous systems (AS) collected by the CAIDA
  project (retrieved from \url{http://www.caida.org}). It
  corresponds to a time-stamped sequence of $E=500,106$ directed
  connections between AS pairs, with a total of $N=53,387$ recorded AS
  nodes. The time-stamps correspond to the first time the connection was
  seen.}

  \item[APS citations]{This dataset contains $E=4,262,443$ time-stamped
  citations between $N=425,760$ scientific articles published by the
  American Physical Society for a period of over $100$ years (retrieved
  from \url{http://journals.aps.org/datasets}).}

  \item[\texttt{prosper.com} loans]{This dataset corresponds to
    $E=3,394,979$ time-stamped directed loan relationships between
    $N=89,269$ users of the \url{prosper.com} website, which provides a
    peer-to-peer lending system (retrieved from
    \url{http://konect.uni-koblenz.de/networks/prosper-loans}).}

  \item[Chess moves]{This dataset contains $38,365$ online chess games
  collected over the month of July 2015 (retrieved from
  \url{http://ficsgames.org/download.html}). The games were converted
  into a bipartite temporal network where each piece and position
  correspond to different nodes, and a movement in the game corresponds
  to a time-stamped edge of the type piece $\to$ position. The resulting
  temporal network consists of $N=76$ nodes and $E=3,130,166$ edges.}

  \item[Hospital contacts]{This dataset corresponds to a temporal
  proximity measurement of patients and health care workers in the
  geriatric unit of an university
  hospital~\cite{vanhems_estimating_2013}. A total of $N=75$ individuals
  were voluntarily tracked for a period of four days, generating
  $E=32,424$ proximity events between pairs of individuals.}

  \item[Infectious sociopatterns]{This dataset corresponds to a temporal
  proximity measurement of attendants at a museum
  exhibition~\cite{isella_whats_2011}. A total of $N=10,972$
  participants were voluntarily tracked for a period of three months,
  generating $E=415,912$ proximity events between pairs of individuals.}

  \item[Reality mining]{This dataset corresponds to a temporal proximity
  measurement of university students and
  faculty~\cite{eagle_reality_2006}. A total of $N=96$ people were
  voluntarily tracked for a period of an entire academic year,
  generating $E=1,086,404$ proximity events between pairs of
  individuals.}

  \item[Beethoven's fifth symphony] A piano
  reduction of Beethoven's fifth symphony, extracted in MIDI format from
  the Mutopia project at \url{http://www.mutopiaproject.org}, represented
  as a sequence of $E=4,223$ notes of an alphabet of size $N=63$.
\end{description}

\section*{Acknowledgements}
We thank M.\ Buongermino Pereira and L.\ Bohlin for comments on the
manuscript. M.R.\ was supported by the Swedish Research Council grant
2016-00796.

\section*{Author contributions}
Both authors conceived the project, performed the research, and wrote the paper. T.P.\ performed the analytical and numerical calculations.

\section*{Competing financial interests}

The authors declare no competing financial interests.

\end{document}